\documentclass[a4paper,11pt]{llncs}

\usepackage{ucs}
\usepackage[utf8x]{inputenc}
\usepackage[T1]{fontenc}

\usepackage{amssymb}
\usepackage{amsmath}
\usepackage{amsfonts}
\usepackage{color,proof}
\usepackage{marginnote}
\usepackage{proof}
\usepackage{listings}

\usepackage{pdflscape}

\title{Detecting Data Races on OpenCL Kernels with Symbolic Execution\thanks{This work was supported by the EU FP7 project CARP (project number 287767).}}
\author{Dino Distefano \and Jeremy Dubreil}
\institute{Monoidics Ltd}
\date{}

\usepackage{xspace}
\usepackage{tikz}

\definecolor{mygreen}{rgb}{0,0.6,0}

\usetikzlibrary{matrix,arrows,automata,positioning}

\tikzset{
    initial text={},
    every state/.style={minimum size=.4cm,draw=blue!50,fill=blue!20},
    secret/.style={circle, draw},
    join/.style={minimum size=.4cm,draw=red!50,very thick,fill=red!20,rectangle},
    whitebox/.style={rectangle, draw=black, minimum width=1cm, minimum height=1cm, rectangle, rounded corners, thick},
    mutebox/.style={rectangle, draw=white, minimum width=1cm, minimum height=1cm, rectangle, rounded corners, thick},
    badbox/.style={rectangle, minimum width=1.4cm, minimum height=1cm, draw=red!50,fill=red!20,very thick},
    goodbox/.style={rectangle, minimum width=1.4cm, minimum height=1cm, draw=mygreen!50,fill=mygreen!20,very thick},
    formula/.style={rectangle, minimum width=1.4cm, minimum height=0.8cm, draw=gray},
    }

\newcommand{\tid}{\texttt{tid}}
\newcommand{\size}{\texttt{size}\xspace}
\newcommand{\start}{\texttt{start}}
\newcommand{\exit}{\texttt{exit}}

\newcommand{\barrier}{\texttt{barrier}\xspace}
\newcommand{\assume}{\texttt{assume}\xspace}

\newcommand{\wait}{\boxdot}

\newcommand{\lvars}{\mathit{LVar}}
\newcommand{\locs}{\mathit{Locs}}
\newcommand{\vals}{\mathit{Vals}}
\newcommand{\heaps}{\mathit{Heaps}}
\newcommand{\stacks}{\mathit{Stacks}}

\newcommand{\sstates}{\mathit{SStates}}
\newcommand{\tstates}{\mathit{TStates}}
\newcommand{\gstates}{\mathit{GStates}}
\newcommand{\vars}{\mathit{PVar}}
\newcommand{\gvars}{\mathit{GVar}}
\newcommand{\defeq}{\stackrel{\mbox{\tiny \it def}}{=}}
\newcommand{\fin}{\mathsf{fin}}
\newcommand{\psto}{{\,{\mapsto}\,}}

\newcommand{\sem}[1]{[ \! | #1 | \! ]}

\newcounter{note_number}

\begin{document}
\maketitle

\begin{abstract}
We present an automatic analysis technique for checking data races on OpenCL kernels. 
Our method defines symbolic execution techniques based on separation logic with suitable abstractions to automatically detect {\em non-benign} racy behaviours on kernels.

\end{abstract}

\section{Introduction}
Graphics Processing Units (GPU) are becoming a popular way to accelerate general-purpose applications.
OpenCL (Open Computing Language) is a cross-platform parallel programming language, developed by the KRONOS group\footnote{\tt http://www.khronos.org}, that is becoming popular for programming GPUs.
OpenCL is a rather low-level language and developing correct parallel applications is challenging and error-prone.

OpenCL programs are composed by two parts: the first one 
called the \textit{host} is
executing on the CPU;  the second part  is a set of functions, called \textit{kernels}, executing on specialised devices like GPUs.

In this paper we develop techniques to automatically analyze OpenCL kernels and discover possible  data races.
In doing so, we focus on distinguishing between {\em benign} and {\em non-benign} races. 
Here, for benign race we intend the possibility for more than one thread to write in the same memory cell the same value. Technically this is a race, however 
it will not result in a non-determinist behaviour of the program as 
the threads write the same value. 
OpenCL programmers explicitly use benign data races as they are harmless. 
An analysis that would flag data races ragardless whether benign or not would give way too many false alarms and would hardly be used in practice.
  
Our method is based on symbolic execution~\cite{BCO05} for computing an overapproximation of the possible behaviours of the program.
More specifically, we develop a new symbolic execution based on separation logic~\cite{REYNOLDS02} tailored to the specifics of OpenCL.
This is challenging task as it requires a special way to deal with the SIMD (Single Instruction Multiple Data) model used by GPUs.



In summary this paper makes the following contribution:
\begin{itemize}
\item We define an operational semantics for SIMD model able to detect non-benign data race.
\item We define a extension of separation logic suitable for the analysis of the SIMD model.
\item We define a new symbolic execution based on separation logic for the SIMD model able to detect non-benign data race.
\end{itemize}
The paper is organised as follows.
In Section~\ref{sec:background} we give a summary of the problems related with kernel and data race.
In Section~\ref{sec:progr-language} we define a minimal OpenCL language.
In Section~\ref{sec:concrete-semantics} we describe a concrete semantics for our language.
In Section~\ref{sec:symbolic-heaps} we define a subset of separation logic for SIMD.
In Section~\ref{sec:symbolic-execution} we describe the symbolic execution.
In Section~\ref{sec:related-work} we discuss other work in the literature. 
Section~\ref{sec:conclusion} concludes the paper.

\section{OpenCL Kernels}
\label{sec:background}
OpenCL Kernels are not full program, but functions embedded in the main program (called the host), and that will be executed on the GPU.
Kernels are executing following the Single Instruction Multiple Data (SIMD) model.
In this model when a kernel is launched by the host, a number of threads (also called {\em work-items}) are executing the same code but on different data.
Threads are grouped in work-groups and both threads and work-groups have a unique identifier that can be referred from the kernel. 
Threads can also query the size of the work-group they belong to, i.e. the number of threads executing the same code in parallel.

\begin{example}
The program in Figure~\ref{fig:kernel} is an example of OpenCL kernel. The special variable {\tt tid} stands for {\em thread id}, the unique identifier of the thread.
Two distinguished threads will only differ from the runtime value of their {\em thread id} which will allow them to manipulate different data.
The {\tt barrier} command synchronises the memory for all the threads executing the kernel. When a thread executes {\tt barrier} it waits until all the other threads have done the same. At that point  the computation resumes.
\end{example}\begin{figure}[t]
   \centering
\begin{verbatim}
  kernel( int A[], int B[], int R[]) { 
   
      R[tid] = A[tid-1] + B[tid+1];
      barrier;
      R[tid] = 2 * R[tid + 1];
  }
\end{verbatim}
 \caption{Example kernel}
 \label{fig:kernel}
\end{figure}

\section{Kernel Programming Language}
\label{sec:progr-language}
We consider the following language for kernels.
It is presented as a control flow
graph consisting of instruction nodes with edges from the nodes to their successors.
Let $\vars$ be a set of program variables.
The set $Comm$ of commands is defined by the following grammar.
\[
\begin{array}{rcl}
	e & ::= & c \mid v \mid op (\stackrel{\rightarrow}{e}) \mid \tid \mid size(v) \\
	b & ::= & e  < e \mid e= e \mid  b \wedge b \mid  \neg b  \\
	Comm & ::= & v := e \mid a[v] := e \mid v := a[v] \\
	  &     & \mid \barrier \mid \texttt{assume}(b) \mid \texttt{assert}(b)
\end{array}
\]
An expression can be a constants $c$, a variable $v$ ($\in \vars$), the special variable $\tid$ for the thread id, the size of an array, or an expression constructed with n-ary operations (e.g. $+,-,*, cos, sqrt$, etc.).
Boolean expression $b$ are standard.
The commands have assignment to a variable or to an element of an array.
The special command $\barrier$ is used in kernels to syncronise threads.
Its informal semantics is the following: when a thread reaches $\barrier$ waits until all threads have reached the same instruction $\barrier$. At that point, threads resume their computation.\\

\newcommand{\Node}{\mathsf{CFGNode}}
\newcommand{\locals}{\mathit{Locals}}
\newcommand{\args}{\mathit{Args}}

\noindent
A kernel is a tuple: $$K(\args) = (\tid, C, \rightarrow, \locals)$$
where the control flow is given as a relation $\rightarrow\ \subseteq \Node \times \Node$. The set of control flow graph nodes are given by $\Node=C \cup \{ \start,\exit \}$ with $C \subseteq Comm$. The kernel begins at the
unique node $\start$ and terminate at the unique $\exit$ node.
A kernel has a unique identifier $\tid$ and a set of argument $\args$  passed by the host program and a set of private variables $\locals$.


\section{Concrete Semantics}
\label{sec:concrete-semantics}
Let $\lvars$ be a set of logical
variables, disjoint from program variables $\vars$, to be used in the
assertion language. Let $\gvars \subseteq \vars$ be a set of global shared variables. 
Let $\locs$ be a countable set of
locations, and let $\vals$ be a set of values that includes
$\locs$, i.e., $\locs \subseteq \vals$. The storage model is given by:
\[
\begin{array}{c}
  \heaps \;\defeq\; \locs \rightharpoonup_\fin \vals
\qquad
  \stacks \;\defeq\; (\vars \cup \lvars) \rightarrow \vals
\\
  \sstates \;\defeq\; \stacks \times \heaps
  \\
   \tstates \;\defeq\; \stacks \times \heaps \times N 
   \\
      \gstates \;\defeq\; \tstates^n \times \sstates  
\end{array}
\]
where $\rightharpoonup_\fin$ denotes a finite partial map. 
 $\sstates$ is the set of  {\em shared states} and they   
  are modelled by a pair $(s,h)$ of a stack and a heap.
$\tstates$ is the set of 
{\em thread states}. They are modelled by a tuple $(s,h,i)$ of a stack, a heap and a thread
identifier. We write $\tstates[i]$ for all the thread states of thread $i$, i.e., $\tstates[i]= \stacks \times \heaps \times \{i\}$.
$\gstates$ is the set of global states (i.e., all thread local states and global states) of the system. A global state has the form:
\[
  (s_1,h_1,1),\cdots,(s_n,h_n,n),\ (s,h)
\]
and includes information for the all the local state of the threads and the shared state of the system.

The concrete semantics is now defined with two transition relations, one reflecting the local computation of a thread (thread-local semantics) and the other describing the global computation of the system (global semantics).

\paragraph{Thread-local semantics.}
The concrete operational semantics of the execution of a thread is defined by a transition relation: 
\[
\leadsto_t  \ \subseteq \  (Comm \times \tstates \times \sstates) \times (\tstates \times \sstates) \cup \{ \bot,\top, \wait \} 
\]
Hence, configurations of our semantics are of the kind: 
\[
  C,\tau,\sigma \leadsto \tau',\sigma' \mbox{ ~or~ } C,\tau,\sigma \leadsto \bot \mbox{ ~or~ } C,\tau,\sigma \leadsto \top \mbox{ ~or~ } C,\tau,\sigma \leadsto \tau \wait \sigma
\] describing the effect of a thread executing the command $C$ in the private state $\tau$ and shared global state $\sigma$.
The result of a step of a transition can be either a modified state $\tau',\sigma'$ or 
one of the special states $\bot$, or $\top$,  or $\wait$.
The special state $\bot$ denotes a runtime error (e.g., a race, an access out of bounds) or that the execution is aborted. $\top$ denotes that a path on a computation does not have continuation.
The special state $\tau \wait \sigma$ indicated that the thread has suspended on the current state $\tau,\sigma$.

In the following we indicate $\tau_s$ and $\tau_h$ for the stack and the heap part of $\tau$. Similarly $\sigma_s$ and $\sigma_h$ denote the stack and the heap of $\sigma$, respectively. For a function $f$ we indicate with $f[v \mapsto n]$ the function update where $v$ now maps to $n$. More precisely:
\[
   f[v \mapsto n](x) = \left\{\begin{array}{ll}
   	n & \quad \mbox{if $x=v$ } \\
	f(x) & \quad \mbox{otherwise}
	\end{array}\right.
\]

The transition relation is defined by the following rules:
\[
\frac{v \notin \gvars \quad n=\sem{e}_{\tau,\sigma}}{v:=e,\tau,\sigma \leadsto_t \tau_s[v \mapsto n],\sigma}
\]
\[
\frac{v \in \gvars \quad n=\sem{e}_{\tau,\sigma}}{v:=e,\tau,\sigma \leadsto_t \tau, \sigma_s[v \mapsto n]}
\]
\[
\frac{a \notin \gvars \quad n=\sem{e}_{\tau,\sigma} \qquad m=\sem{v}_{\tau,\sigma}}{a[v]:=e,\tau,\sigma \leadsto_t \tau_h[\tau_s(a)+m \mapsto n],\sigma}
\]
\[
\frac{a \in \gvars \quad n=\sem{e}_{\tau,\sigma} \qquad m=\sem{v}_{\tau,\sigma}}{a[v]:=e,\tau,\sigma \leadsto_t \tau, \sigma_h[\sigma_s(a)+m \mapsto n]}
\]
\[
\frac{v \notin \gvars \quad n=\sem{w}_{\tau,\sigma} }{v:=a[w],\tau,\sigma \leadsto_t \tau_s[\tau_s(v) \mapsto \tau_h(\tau_s(a)+n)],\sigma}
\]
\[
\frac{v \in \gvars \quad n=\sem{w}_{\tau,\sigma} }{v:=a[w],\tau,\sigma \leadsto_t \tau, \sigma_s[\sigma_s(v) \mapsto \sigma_h(\sigma_s(a)+n)],\sigma}
\]
\[
\frac{\sem{b}_{\tau,\sigma}=false }{\texttt{assert}(b),\tau,\sigma \leadsto_t \bot}
\]
\[
\frac{\sem{b}_{\tau,\sigma}=true }{\texttt{assert}(b),\tau,\sigma \leadsto_t \tau,\sigma}
\]
\[
\frac{\sem{b}_{\tau,\sigma}=false }{\texttt{assume}(b),\tau,\sigma \leadsto_t \top}
\]
\[
\frac{\sem{b}_{\tau,\sigma}=true }{\texttt{assume}(b),\tau,\sigma \leadsto_t \tau,\sigma}
\]
\[
\frac{}{\texttt{barrier},\tau,\sigma \leadsto_t \tau \wait \sigma}
\]
The barrier rule expresses  that a thread executing a barrier suspend its computation in the current local and shared state (indicated by $\tau \wait \sigma$).

\paragraph{Global Semantics.}
The global semantics is defined with a transition relations
\[
 \leadsto_g \ \subseteq (Comm \times \gstates ) \times \gstates
\]
In the global semantics we have the following rules. The first two rules give the interleaving of kernels operations. If a thread makes a local step, this is observed in the global 
execution of the system. 
\[
\frac{C,\tau[i],\sigma \leadsto_t \tau'[i],\sigma'}{C,T,\sigma \leadsto_g T[i \mapsto \tau'[i]],\sigma'}
\]
The second rule records the special case when a thread
waits at a barrier

\[
\frac{C,\tau[i],\sigma \leadsto_t \tau[i] \wait \sigma}{C,T,\sigma \leadsto_g T[i \mapsto \tau[i] \wait \sigma],\sigma}
\]
The next rule describes the case where all the threads have reached the barrier and are in a waiting state. If the shared state $\sigma$ is the same for every threads, then no races have occurred. The rule removes the waiting state allowing threads to resume their computation.
\[
  \frac{}{(\tau_1 \wait \sigma, \dots, \tau_n \wait\sigma), \sigma \leadsto_g (\tau_1,\dots,\tau_n),\sigma}
\]
Finally, the last rule detects if a race has occurred. This is identified when there exists a thread $i$ that is waiting in a shared state $\sigma_i$ which is different from the global shared state $\sigma$. The global shared state is  obtained when the last thread in the particular schedule considered entered in the waiting state.  In the case the system has a race and  the global computation reaches the error state.
\[
  \frac{\exists i. \sigma_i \neq \sigma}{(\tau_1 \wait \sigma_1, \dots, \tau_n \wait\sigma_n), \sigma \leadsto_g \bot}
\]
Notice that these rules describe any scheduling for threads. Moreover, notice that the race rule reaches the error state only for {\em non-benign} races. 
In case of benign races (i.e., more than one thread writing the same variable or memory location with the same value) the computation continues normally. 
\begin{definition}
A command $C$ has a {\em race} if there exists a thread state $T$ and a shared state $\sigma$ such that $C,T,\sigma \leadsto^*_g \bot$.
\end{definition}
According to our operational semantics, a program is racy if it presents non-deterministic behaviour on the shared state when the threads syncronise by means of a barrier. This includes also the last implicit barrier at the end of the kernel.

\section{Symbolic Heaps}
\label{sec:symbolic-heaps}
Symbolically, we use formulae in separation logic to represent set of states
during the computation of the kernel.
The subset of formulae of separation logic we use in this paper is an extension of symbolic heaps~\cite{DistefanoOHearnYang06}. We also make specific considerations about arrays as they are fundamental data structures in OpenCL.\\

\noindent
Arrays are encoded in separation logic as follows:
\begin{itemize}
 \item we denote by $v[i]$ the value at address $v + i$
 \item we denote by $x \psto A[i, j]$ an array segment, that is:
 	$$x \psto A[i, j] \Leftrightarrow  (x + i \psto -) * (x + i + 1 \psto -) * \cdots * (x + j \psto -)$$
\end{itemize}

\noindent
Symbolic heaps are defined by the following grammar:
\[
\begin{array}{lcl}
	E & ::= & c \mid v \mid v' \mid \size(v) \mid E \ op \ E \\
	\Pi & ::= & E = E \mid E \neq E \mid E \leq E \mid F \mid true \mid \Pi \wedge \Pi \\
	F & ::= & \lambda i . E \mid \eta(\Pi, F, F) \\
	S & ::= & E \psto E \mid E \psto A[n, m | F] \\
	\Sigma & ::= & S \mid emp \mid \Sigma * \Sigma \\
	H & ::= & \exists x' .(\Pi \wedge \Sigma)
\end{array}
\]
Primed variables are implicitly existentially quantified.
The other expression in $\Pi$ are standard.
The points-to predicate $E \psto E'$ is  standard and denotes an allocated cell at address $E$ with content $E'$.
The predicate $E \psto A[n, m | F]$ is used to encode arrays. Its  meaning is  that $E$ points-to an array segment with index ranging from $n$ to $m$ and values defined by the function expression $F$. 
The latter, in the simple case, is a function $\lambda i.E$ that given an index $i$ returns the value of the array at that index\footnote{Note that $E$ may depend on $i$}.
Alternatively, function expressions $F$ can be defined by successions of updates using the constructor $\eta$:
\[
 	\begin{array}{lcll}
	\eta(\phi, g, f) = & \lambda i & . &
		\left\{
			\begin{array}{l}
					g(i) \text{ when $\phi(i)$ is true}\\
					f(i) \text{ otherwise}
			\end{array}
		\right.
	\end{array}
\]
%
Therefore $\eta(\phi,g,f)$ describe the content of the array to be $g(i)$ for the subset of index which make $\phi(i)$ true otherwise $f(i)$.
This expression allows us to describe updates of the array done simultaneously by several threads\footnote{Note that we do not check here if such updates introduces data-races. This will be done in section~\ref{sec:symbolic-execution}}.\\

\noindent
In the sequel, we will use the following notation:

$$a \psto F ~\text{ to mean }~ a \psto A[(0, size(a) - 1 | F]$$
when an array contains $F$ in all its elements.

\paragraph{Semantics in Thread state. }
The semantics of the fragment of separation logic we use here is defined on thread states. The definition is divided in three parts. We first give the semantics $\models_P$ for the pure part of the fragment. 
\[\begin{array}{lclll}
s,h,i & \models_P & E_1 = E_2  & \mbox{ iff } & s(E_1) = s(E_2)\\
s,h,i & \models_P & E_1 \neq E_2 & \mbox{ iff } & s(E_1) \neq s(E_2) \\
s,h,i & \models_P & E_1 \leq E_2 & \mbox{ iff } & s(E_1) \leq s(E_2) \\ 
s,h,i & \models_P & true & \mbox{ iff } & \mbox{always }\\
s,h,i & \models_P & \Pi_1 \wedge \Pi_2  & \mbox{ iff } &  s,h,i \models_P \Pi_1 \mbox{ and } s,h,i\models_P \Pi_2 \\
\end{array}\]
We then define the semantics $\sem{\cdot}: F \rightarrow \tstates \rightarrow N \rightarrow  \vals$ of functions defining the content of arrays. It is defined in terms of $\models_P$ and the semantics of pure expressions $\sem{\cdot}_s$ as follows:
\[
\begin{array}{lcl}
\sem{\lambda j. E}_{s,h,i} & = & \sem{E(j)}_s 
\\[2ex]
\sem{\lambda j . \eta(\Pi, F_1, F_2)}_{s,h,i} & = & \left\{\begin{array}{ll}
	\sem{F_1(j)}_{s,h,i} & \mbox{ if $s,h,i \models_P \Pi$} 
	\\
	\sem{F_2(j)}_{s,h,i} & \mbox{ otherwise}
	\\
	\end{array}
	\right.
\end{array}
\]
Next we give the satisfaction relation for the spatial part of the formulae $\models_S$:
\[\begin{array}{lclll}
s,h,i & \models_S & E_1 \psto E_2 & \mbox{ iff } &  dom(h)=\{ s(E_1)\} \mbox{ and } h(s(E_1))=s(E_2)\\ 
s,h,i & \models & E_1 \psto A[n, m | F]  & \mbox{ iff } & dom(h)=\{s(E_1)+n,\dots,s(E_1)+m \} 
\\ & & & & \mbox{and } n\leq j \leq m: h(s(E_1)+j)=\sem{F(j)}_{s,h,i} \\
s,h,i & \models_S & emp & \mbox{ iff } & dom(h)=\emptyset\\ 
s,h,i & \models_S & \Sigma_1 * \Sigma_2 & \mbox{ iff } & \exists h_1,h_2{:}  \ dom(h_1) \cap dom(h_2) = \emptyset, h_1 \cup h_2 = h \mbox{ and} \\
& &   & &  s,h_1,i \models_S \Sigma_1 \ \mbox{ and } \ s,h_2,i \models_S \Sigma_2 
\end{array}\]
Finally the semantics $\models$ of a complete symbolic heap is defined as:
\begin{eqnarray*}
s,h,i  \models_S  \exists \vec {x'} .(\Pi \wedge \Sigma) & \mbox{ iff } & \exists \vec v{:} \ 
s,h,i \models_P \Pi[\vec v/\vec {x'}]  \mbox{ and } s,h,i \models_S \Sigma[\vec v / \vec{x'}]
\end{eqnarray*}

\section{Symbolic Execution}
\label{sec:symbolic-execution}
In this section we define the symbolic execution for OpenCL kernels. It allows to 
automatically build over-approximations of kernel behaviours. During the computation  of the over-approximations the symbolic execution can detect the possibility 
of a non-benign race and flag it.

One difficulty in defining the rules of symbolic execution
is to formalise the ability for kernels to use the special variable $\tid$ when dereferencing shared data structures (e.g., arrays).
Because the execution semantics of kernels is Single Instruction Multiple Data (SIMD), an update to an array using $\tid$ involve the update of the array in multiple locations. These locations depend on the possible value $\tid$ can range at the program point where the update happens.
In general, there are no constraints on $\tid$, then a single update actually
translates to an update for each kernel.
Hence $\tid$ has to be treated differently in the rules.
Its interpretation in the formulae corresponds to one of an universally quantified variable.


%
\begin{table}[t]
\begin{center}
\begin{tabular}{c}
\\[2ex] 
$\infer{~ v := e,~ \Pi \wedge \Sigma \implies v = e[v'/v] \wedge (\Pi \wedge \Sigma)[v'/v]}{}$
\\[2ex] 
$\infer{v: = a[e],~ \Pi \wedge \Sigma * a \psto f \implies v = f(e[v'/v]) \wedge (\Pi \wedge \Sigma * a \psto f)[v'/v] }{\Pi \wedge \Sigma \vdash 0 \leq e < \texttt{size}(a)}$
\\[2ex]
$\infer[\mbox{where }\phi{:} \ i  \mapsto  \Pi \wedge i = e]{a[e] := e',~ \Pi \wedge \Sigma * a \psto f \implies \Pi \wedge g = \eta(\phi, f, e') \wedge \Sigma * a \psto g }{\Pi \wedge \Sigma \vdash 0 \leq e < \texttt{size}(a)}$ 
\\[2ex]
\infer{C(a[e]),~ \Pi \wedge \Sigma \implies \bot}{\Pi \wedge \Sigma \nvdash 0 \leq e < \texttt{size}(a)}
\\[2ex]
\infer{\assume(b),~ \Pi \wedge \Sigma \implies (\Pi \wedge b) \wedge \Sigma}{(\Pi \wedge b) \wedge \Sigma \nvdash false}	
\\[2ex]
\infer{\assume(b),~ \Pi \wedge \Sigma \implies \top}{(\Pi \wedge b) \wedge \Sigma \vdash false}	
\\[2ex]
\infer{\texttt{assert}(b),~ \Pi \wedge \Sigma \implies \Pi \wedge \Sigma}{ \Pi \wedge \Sigma \vdash b}	
\\[2ex]
\infer{\texttt{assert}(b),~ \Pi \wedge \Sigma \implies \bot}{ \Pi \wedge \Sigma \nvdash b}	
\end{tabular}
\end{center}
\caption{Symbolic Execution Rules}
\label{tab:symbolic-execution}
\end{table}%

%
%
%

 
 The symbolic execution is defined for the language in Section~\ref{sec:progr-language}. 
 Our semantics assume to have a theorem prover not necessarily complete on entailments of the following kind:
 \[
	\Pi \wedge \Sigma \vdash \Pi' \wedge \Sigma'
 \]
The definition of this prover is out of the scope of this paper.
 
The rules are defined in Table~\ref{tab:symbolic-execution} and they symbolically describe the behaviour of a thread by transforming formulas $\Pi\wedge\Sigma$ according to the effect of a command $C$. These rules concern a single symbolic heap. However to increase the precision of the analysis in presence of loops we 
admit sets of symbolic heaps on each node in the control flow-graph.
We lift the symbolic semantics of threads $\implies$ to sets 
 $\Longrightarrow_s$. In this semantics symbolic states are sets of symbolic heaps.
 
 Given an abstract state represented as a set of symbolic heaps $\mathcal{H}$, and a command $C \neq \barrier$ the symbolic transition relation is defined by:
%
\[\begin{array}{ccc}
	\infer{C,~ \mathcal{H} \Longrightarrow_s  {\cal H'}}{ {\cal H'}=\{ H' \mid \exists H \in \mathcal{H},~ C, H \implies H' \}  \qquad \bot \notin {\cal H'}}
	& \qquad &
		\infer{C,~ \mathcal{H} \Longrightarrow_s  \bot}{   \exists H \in \mathcal{H}: ~ C, H \implies \bot }
\end{array}
\]
According to this rules, if the execution of the command on one symbolic heap results in an illegal action (e.g., an array out-of-bound), then the global system goes to error.

Notice that the semantics $\Longrightarrow_s$ represents an over-approximation of the execution of a thread. The resulting collective semantics is parameterised on $\tid$. In other words, $\Longrightarrow_s$ represents the behaviour of a generic threads. The behaviour of a specific thread is then obtained instantiating $\tid$ to a
specific value. Therefore, to observe data-races in the system we can instantiate the parametric symbolic execution to two distinct threads instances. This is the rationale of the rule below for the $\barrier$ statement. 
Notice moreover, that this allows us to perform the analysis of the parameterised kernel as if it was a sequential process, therefore avoiding to consider all possible interleaving and consequently state-space explosion.

To define the rule for $\barrier$ we first need some auxiliaries definitions.
Given a symbolic heap $H$, the function $rename(i, H)$ replaces all occurrences of $\tid$ by $i$, and renames all private variables of $H$ to names that are unique to thread $i$.
The objective of $rename$ is that for two thread id variables $i \neq j$, there is no name conflicts on private variables between $rename(i, H)$ and  $rename(j, H)$.

Given two symbolic heaps $H_1$ and $H_2$, and two expressions $e_1 \in H_1$ and $e_2 \in H_2$, we consider given two functions: 
\[
\begin{array}{l} 
compare((e_1, H_1), (e_2, H_2))
\\[2ex]
disjoint((e_1, H_1), (e_2, H_2))
\end{array}
\] 
$\mathit{compare}$ tries to prove that the expressions $e_1$ and $e_2$ denote the same value in both symbolic heaps.
$\mathit{disjount}$ tries to prove that expressions $e_1$ and $e_2$ denotes different values. These functions return $\mathit{false}$ when it cannot be stated from the formula that the two expressions $e_1$ and $e_2$ denote the same value ($\mathit{compare}$) or different ($\mathit{disjoint}$).

\begin{definition}
\label{def:no_benign_race}
Let $\mathcal{H}$ be a symbolic state. We say that ${\cal H}$  is (data) {\em race-free}, 
 written $\mathsf{NoRace}(\mathcal{H})$, when:
 \[
 	\begin{array}{c}
 	\forall i,j,~i \neq j, \quad \forall H, H' \in \mathcal{H} \\ 
	\tilde{H} = rename(i, H) \text{ and } \tilde{H}' = rename(j, H'), \\
	\forall e_1 \psto e_2 \in \tilde{H}, \ \forall e_1' \psto e_2' \in \tilde{H}', \\
 \neg disjoint((e_1, \tilde{H}), (e_1', \tilde{H}')) \implies compare((e_2, \tilde{H}), (e_2', \tilde{H}')) 
	\end{array}
 \]
\end{definition}
This definition states that taking any pair of symbolic heaps $H$, $H'$ in $\mathcal{H}$ instantiated to two arbitrary distinct instances of $\tid$ modelling two different threads, there is no data race whenever the shared state has the same value.
More precisely, we need to check that for all the memory cells $e_1$, $e_1'$ in the shared state for which we cannot prove that they represent different addresses, we should be able to prove that the values they point to are the same. 
This definition reflects our intuitive notion of benign race. We flag a potential race only if we cannot prove that the share state of all threads at barrier time is deterministic.
Notice that when checking for disjointness we require that we should be able to prove  in both $\tilde H$ and $\tilde{H'}$ that the left-hand side of the points-to $e_1$ and $e_1'$ are different. Otherwise, we require that we can prove in both heaps that $e_2$ and $e_2'$ are equal. Failing that our definition should declare that there might be a potential race.

Based on this, we define the abstract rule for \barrier as follows:
\[
\begin{array}{ccc}
	\infer
		{ 	\barrier,~ \mathcal{H} \Longrightarrow_s \mathcal{H}
		 }
		{ \mathsf{NoRace}(\mathcal{H}) } 
		& \quad  \quad &
			\infer
		{ 	\barrier,~ \mathcal{H} \Longrightarrow_s \bot
		 }
		{ \neg \mathsf{NoRace}(\mathcal{H}) }
\end{array}
\]
Notice that in the reading of the precondition of the second  rule should be intended to mean when is it not possible to prove $\mathsf{NoRace}(\mathcal{H})$.

\subsection{Discussion}
\label{discussion}
Although the symbolic execution described in this section will be able to detect interesting data races on most concrete examples, the analysis is not sound. Indeed, there can be cases where the analysis will fail to detect branches depending on the values of arrays. For example, the code sample of Figure~\ref{fig:counterexample}, the analysis will not be able to detect the race of the global variable $g$.
\begin{figure}[t]
   \centering
\begin{verbatim}
  kernel( int A[]) { 
    g = 0
    A[tid] = 0
    barrier
    A[tid] = 1
    if A[tid + 1] = 0 then g = 1
    barrier
  }
\end{verbatim}
 \caption{Counter example}
 \label{fig:counterexample}
\end{figure}
Indeed, g can in practice have the value 0 or the value 1. But, according to the rules for the symbolic execution, the array $A$ will get the value $A \mapsto \lambda i. 0$. Then, the symbolic execution will consider the test \texttt{A[tid + 1] == 0} as always false, and forget the branch where $g$ gets updated to $1$ and miss the data race. It seems that this issue can be fixed by just keeping the same rules, but considering the interleaving product of two threads with two distinct thread id $i$ and $j$ such that $i < j$. The predicate {\sf NoRace} does not need to be updated as these two arbitrary threads $i$ and $j$ will be the same than the ones present in Definition~\ref{def:no_benign_race}. 
We intend to investigate the possible extensions to a sound approach in the future.
The soundness should be proved against the concrete semantics defined in Section~\ref{sec:concrete-semantics}.

\section{Related Works}
\label{sec:related-work}

Recently several approaches to formal analysis of GPU kernels have
been
proposed~\cite{PUG,BettsCDQT_OOPSLA2012,CollingbourneDKQ_ESOP2013,DBLP:conf/ppopp/LiLSGGR12,KLEECL,PLDI2012,BYTECODE}.

In~\cite{BettsCDQT_OOPSLA2012}, the authors present an approach to verify intra-group data races and barrier divergence. Their approach is based on modelling the OpenCL kernels semantics using predicated execution and a synchronous, delayed visibility semantics to deal with the modifications of shared memory. With this delayed visibility semantics, similar to transactional memory, every thread updates a local copy of shared variables, and the consistency on these modifications are verified at the barrier level. At these synchronisation points, it must be true that no shared variable modified by one thread has been read or modified by another thread in the delayed semantics. Practically, this theory is implemented by translating OpenCL programs into Boogie~\cite{Boogie} and use the Z3~\cite{Z3} theorem prover to prove consistency between delayed reads on writes. Instead, in our approach, we consider a more abstract definition of race freedom based on the absence of non-determinism at the barrier level. We also use a crafted logic based on separation logic to define the abstract states of our symbolic execution. Unlike approaches based on theorem prover, our approach aims at being fully automatic.  The technique of~\cite{PUG} shares similarities with~\cite{BettsCDQT_OOPSLA2012}: it is also based on generating constraints to be solved by a theorem prover.

Another line of work that is worth comparing with our approach is the tool GKLEE~\cite{DBLP:conf/ppopp/LiLSGGR12}.
In this paper, the author present a technique based on the existing theory behind the tool KLEE~\cite{klee} which uses static analysis techniques to generate test cases, and evaluate the test coverage provided by a set of test cases.
The approach of~\cite{DBLP:conf/ppopp/LiLSGGR12} is based on defining a symbolic virtual machine that will accurately model the behaviours of kernel threads. Based on this, the tool generates sets of tests to force kernels to execute and detect errors relevant for GPU programming such as data races, but also to outline possible performance issues related to memory access. Unlike~\cite{BettsCDQT_OOPSLA2012}, the approach of~\cite{DBLP:conf/ppopp/LiLSGGR12} does not aim at providing soundness on data races freedom, but provides instead a level of confidence with the concept of test coverage.  The KLEE-CL tool~\cite{KLEECL} is a similar method for analysing GPU kernels, also based on KLEE.

\section{Conclusions and Future Work}
\label{sec:conclusion}
We have introduced a novel automatic approach for the analysis of OpenCL Kernels based on separation logic.
Our method uses a tailored symbolic execution able to detect non-benign races.

In the future we plan to extend our technique from a race detector to a race-free verifier. One avenue we intend to try  discussed in Section~\ref{discussion}.
Moreover, in the future we plan  to  automatically synthesise loop invariants. This requires the design of  new sophisticated finite abstractions for arrays in OpenCL. We will use techniques from abstract interpretation~\cite{CousotCousotPOPL77} 

Another avenue we intend to investigate is the possibility to automatically generate  preconditions such that kernels can be run in a data-race free way. 

An implementation is left for future work.

\bibliographystyle{plain}
\bibliography{biblio}

\end{document}